\documentclass[aps,prb,longbibliography,reprint,groupedaddress,floatfix]{revtex4-1}
\usepackage{amssymb,amsmath}
\usepackage{graphicx}
\usepackage{epsfig}
\usepackage{bm}

\begin{document}
\title{Parallel implementation of electronic structure eigensolver using a
partitioned folded spectrum method}
\author{E.L. Briggs}
\affiliation{Center for High Performance Simulation and Department of Physics,
 North Carolina State University, Raleigh, North Carolina, 27695-7518}
\author{C.T. Kelley} 
\affiliation{Center for High Performance Simulation and Department of Mathematics,
 North Carolina State University, Raleigh, North Carolina, 27695-8205}
\author{J. Bernholc}
\affiliation{Center for High Performance Simulation and Department of Physics,
 North Carolina State University, Raleigh, North Carolina, 27695-8205}
\date{\today}

\begin{abstract}
 A parallel implementation of an eigensolver designed for electronic 
structure calculations is presented. The method is applicable to
computational tasks that solve a sequence of eigenvalue problems where
the solution for a particular iteration is similar but not identical to
the solution from the previous iteration. Such problems occur frequently
when performing electronic structure calculations in which the eigenvectors
are solutions to the Kohn-Sham equations. The eigenvectors are represented
in some type of basis but the problem sizes are normally too large for
direct diagonalization in that basis. Instead a subspace diagonalization
procedure is employed in which matrix elements of the Hamiltonian operator
are generated and the eigenvalues and eigenvectors of the resulting reduced
matrix are obtained using a standard eigensolver from a package such as
LAPACK or SCALAPACK. While this method works well and is widely used, the
standard eigensolvers scale poorly on massively parallel computer systems
for the matrix sizes typical of electronic structure calculations. We
present a new method that utilizes a partitioned folded spectrum scheme
(PFSM) that takes into account the iterative nature of the problem and
performs well on massively parallel systems. Test results for a range of
problems are presented that demonstrate an equivalent level of accuracy
when compared to the standard eigensolvers, while also executing up to an
order of magnitude faster. Unlike O(N) methods, the technique works
equally well for metals and systems with unoccupied orbitals as for
insulators and semiconductors. Timing and accuracy results are presented
for a range of systems, including a 512 atom diamond cell, a cluster of
13 C60 molecules, bulk copper, a 216 atom silicon cell with a vacancy, 
using 40 unoccupied states/atom, and a 4000 atom aluminum supercell.
\end{abstract}
\maketitle

\section{Introduction}
Electronic structure calculations based on density functional theory are 
widely used in physics and materials science. These calculations usually
involve numerical solutions of the Kohn-Sham equations for a set of
eigenvalues and eigenvectors. 

\begin{equation}
\label{eq:kohnsham}
\bm H_{ks} \bm [\psi_i] = -\frac{1}{2} \nabla^2 \psi_i + V_{eff} \psi_i = 
\lambda_i \psi_i \;\;\; i=1,N
\end{equation}

where
\begin{equation}
 \label{eq:veff}
 V_{eff} = V_{hartree} + V_{xc} + V_{ionic}
\end{equation}

Self consistent pseudopotentials\cite{PhysRevB.8.1777, PhysRevLett.43.1494, 
 zunger1978first} of various forms are used to represent 
$V_{ionic}$ while $V_{hartree}$ and $V_{xc}$ are the exchange correlation and
hartree potentials. A typical solution process proceeds by generating an
initial set of electronic orbitals and an associated charge density. A linear
combination of atomic orbitals from the constituent ions is often used
for this together with a superposition of atomic charge densities as a
starting density.
The initial wavefunctions may be represented by expansions in basis
functions, such as plane waves\cite{PhysRevB.12.5575,PhysRevLett.55.2471}, augmented waves
\cite{PhysRevB.41.5414} or grid based discretizations\cite{PhysRevB.54.14362,
chelikowsky1994finite,PhysRevB.76.085108}. The initial density is then used
to compute exchange correlation and hartree potentials using functionals of
the electron density\cite{perdew1996generalized}, which along with the ionic
potentials form $V_{eff}$. This representation of $V_{eff}$ is then
substituted back into the Kohn-Sham equations that are solved for a new set
of electronic orbitals. This process is repeated until self consistency is
reached, where self consistency is defined as the difference between 
$V_{eff}(in)$ and $V_{eff}(out)$ potentials being less than a specified
tolerance.

A variety of methods have been used to solve Eq.(\ref{eq:kohnsham}) for each
self consistent step. Many of these methods include a subspace diagonalization
and solution of a real symmetric eigenvalue problem as part of the
procedure. For large systems this will normally be the most expensive part of
the calculations as the work required scales as the cube of the number of
electronic orbitals. Since this is the main computational bottleneck for large
DFT calculations, considerable work has been carried out to either
eliminate the cubic scaling terms entirely or reduce their prefactor.
Algorithms that fall into the first category often rely on the restricted
spatial range made possible by real space methods. These include both grid
based\cite{PhysRevB.62.1713,PhysRevB.73.115124}, and localized basis function
approaches.\cite{soler2002siesta} The present work, which is a partitioned
folded spectrum\cite{wang1994solving} method (PFSM) falls into the second
category and works by distributing eigenpairs over independent processing
nodes. Eigenpairs on a given node are solved for using submatrix
diagonalizations and an iterative folded spectrum technique is used to couple
the submatrix solutions to the full matrix. No distinction is made between
occupied and unoccupied states, in contrast to the spectrum slicing method
developed by Schofield et~al.\cite{schofield2012spectrum} where only the
occupied eigenspectrum is divided into discrete slices. These are distributed
over the nodes and filtering functions are used to restrict the eigenpairs
on a specific slice to a discrete range of the occupied orbitals. The current
method also shares some similarities with the shift and invert parallel
spectrum transformation (SIPs) used by Zhang 
et~al.\cite{Zhang:2007:SSP:1236463.1236464} for
tight binding calculations, which generate sparse matrix eigenproblems
as opposed to the dense matrices addressed in the current work.
Their approach takes advantage of the sparsity to reduce the scaling factor
from $O(N^3)$ to $O(N^2)$ for sufficiently sparse matrices.

Since direct diagonalization of Eq.(\ref{eq:kohnsham}) is difficult, a 
subspace diagonalization procedure is employed in which matrix elements
of the Hamiltonian operator are generated. The eigenvalue problem for the
resulting reduced matrix is then given by.

\begin{equation}
\label{eq:eigenproblem}
\bm A\bm x_i = \lambda_i \bm  x_i \;\;\; i=1,N
\end{equation}

This can be solved using a variety of standard library routines. These include
LAPACK and/or MAGMA\cite{agullo2009numerical} routines, which utilize a single
processing node, and ScaLapack, which is designed to use multiple nodes. When
running on multiple nodes of a supercomputer or cluster, LAPACK is a poor
choice since it can only utilize the processing power of a single node. On a
system with GPU's, the Magma libraries are a viable alternative for N ranging
up to a few thousand. While Magma can currently only apply the processing
power of a single node, GPU's are well suited for this type of problem and
are so fast that solutions may be achieved in an acceptable amount of time.
For problems where N is greater than a few thousand, ScaLapack usually becomes
the best choice. The exact size of N depends on the relative balance of
computational power to communications speed on the computer system being
used. Unfortunately, for any given value of N a point will be reached where
the performance of ScaLapack scales poorly as the number of nodes is
increased due to communications overhead. Current implementations of ScaLapack
are also unable to effectively utilize GPU's, leaving large amounts of
processing power unused.

The current work is designed to address the shortcomings of these methods.
The main design goal is to minimize the wall clock time required for a
solution of Eq.(\ref{eq:eigenproblem}). Exceptional efficiency in terms of the
total operations count is not necessarily needed to achieve this. Instead, the
focus is on effectively using as much of the processing power available,
as opposed to the current methods that may leave much of that power untapped
on large clusters.

\section{Methodology}

Let $\{\lambda_i\}$ and {$\{\bm u_i\}$} be the eigenvalues and eigenvectors of
$\bm A$. We order $\lambda_i$ in distance from the target $\epsilon$. So

\begin{equation}
 \label{eq:eigs1}
|\lambda_1 - \epsilon| < |\lambda_2 - \epsilon| \le |\lambda_3 - 
\epsilon| \le .... |\lambda_n - \epsilon|.
\end{equation}

The eigenvector corresponding to the eigenvalue closest to $\epsilon$ may be
determined with the iteration.

\begin{equation}
 \label{eq:eigs2}
\bm x^{n+1} = (\bm I - \alpha(\bm A - \epsilon \bm I)^2)\bm x^n
\end{equation}

This is just the power method for the matrix.

\begin{equation}
 \label{eq:eigs3}
\bm I - \alpha(\bm A - \epsilon \bm I)^2
\end{equation}

and it will converge to the desired eigenvalue

\begin{equation}
 \label{eq:eigs4}
1 - \alpha(\lambda_1 - \epsilon)^2 ,
\end{equation}

if the eigenvalues of $\bm I - \alpha(\bm A - \epsilon \bm I)^2$ are
ordered so that for each n

\begin{equation}
 \label{eq:eigs6}
\alpha | \lambda_n - \epsilon| < 1
\end{equation}

and

\begin{equation}
 \label{eq:eigs7}
\alpha <  \frac{1}{| \lambda_n - \epsilon|}.
\end{equation}

Then

\begin{equation}
\label{eq:eigs5}
\begin{split}
|1 - \alpha(\lambda_n - \epsilon)^2| \le |1 - \alpha(\lambda_{n-1} - 
\epsilon)^2| \le ... \\ \le ... |1 - \alpha(\lambda_{2} - 
\epsilon)^2| < |1 - \alpha(\lambda_1 - \epsilon)^2|
\end{split}
\end{equation}

for $1 \le i \le N -1 $.


If (\ref{eq:eigs7}) is satisfied then given an initial estimate of an
eigenvector $\bm x_i^j$ and a shift value $\epsilon$, the iterative process
will produce an eigenvector with an eigenvalue close to $\epsilon$.

\begin{equation}
\label{eq:foldedspectrum}
 \bm x_i^{j+1} = \bm x_i^j - \alpha (\bm A-\epsilon \bm I)^2 
 \bm x_i^j \;\;\;\;0 < \alpha < 1
\end{equation}

This may be implemented via a sequence of matrix-vector products which
eliminates the need to explicitly square the matrix $\bm A-\epsilon \bm I$.
Since the iterative process is only guaranteed to produce eigenvectors for
eigenvalues close to the shift value $\epsilon$, we must also have a way of
generating a suitable set of shifts and initial eigenvectors. We do this by
solving a sequence of eigenvector problems generated from overlapping diagonal
submatrices of the original matrix $\bm A$.

\begin{figure}[ht]
\includegraphics[scale=0.5]{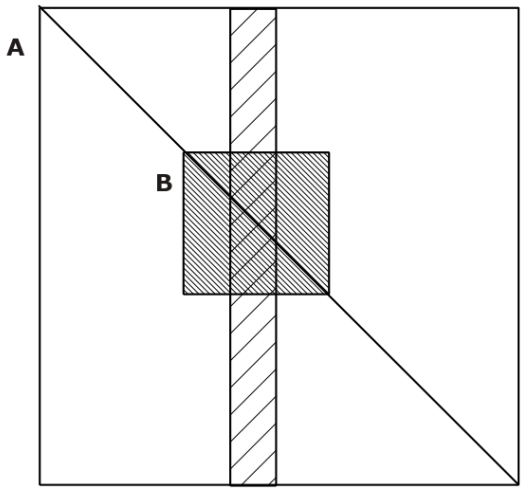}
\caption{The region with coarse grained cross hatching represents
the set of eigenvectors assigned to an individual processing node.
The region with fine grained cross hatching represents both submatrix
$\bm B$ and the set of eigenvectors associated with a solution of 
$\bm B \bm y_i=\lambda \bm y_i$}
\label{submatrices}
\end{figure}

In particular, each diagonal submatrix $\bm B$ is of order M where
$M = \gamma N$ and $0 < \gamma \le 1$. Since the work required to solve each
submatrix scales as $M^3$, each individual submatrix requires $\gamma^3$ less
work. The total computational work required then depends on the magnitude of
$\gamma$ and the number of submatrices needed. Even when this is larger than
the work required by a standard solver such as LAPACK or ScaLapack, the wall
clock time required for a solution can be significantly less because the
algorithm is better adapted for massively parallel computer architectures. By
assigning individual submatrices to smaller blocks of processing nodes, the
scalability constraints found in the standard solvers are alleviated. In our
tests we have found for values of N ranging up to a few thousand (common in
electronic structure calculations), assigning one submatrix per processing
node works well on machines with GPU accelerators. For this case using a
computing platform with P processing nodes, a slice of eigenvectors of size
$m = N/P$ may be assigned to each node as shown in Fig.(\ref{submatrices}).

  The eigenvector solutions $\bm y_i$ for the  submatrices are used to
construct the starting vectors $\bm x_i^0$ for Eq.(\ref{eq:foldedspectrum})
and the corresponding eigenvalues are used as the shifts $\epsilon$. Since
$M < N$, the elements of $\bm x_i$ that do not have a corresponding component
in the submatrix solution are set to zero.  The iterative process from
Eq.(\ref{eq:foldedspectrum}) is then applied and after all eigenvectors have
been processed, Gram-Schmidt orthogonalization is used to ensure
orthonormality of the resulting eigenvectors. Note that since this process
is applied on a subspace and is readily parallelized, it represents only a
moderate fraction of the total work.

This method will not work reliably for an arbitrary matrix. This may be
demonstrated by taking a set of eigenvectors that satisfy 
Eq.(\ref{eq:kohnsham}) and randomly mixing them to generate $\bm A$.
In this case the eigenvectors of the submatrices will consist of linear
combinations of eigenvectors from the full spectrum and the iterative process
described in Eq.(\ref{eq:foldedspectrum}) can produce nearly identical
estimates for the submatrix $\lambda_i$ that will not span the full spectrum.

A key condition for success then is that A be diagonally dominant. We have
generally found that using standard diagonalization for the  first 3 to 5
SCF steps is sufficient when the submatrix widths are 30 percent of the full
width of A and the number of eigenvectors computed via Eq. (11) from each
submatrix is less than 10 percent of the full width. We thus expect major
savings during the usual applications of DFT, such as geometry optimization
and ab initio molecular dynamics, in which only the first ionic step would
require some standard diagonalizations, while PFSM would be used for the
many subsequent steps.

\section{Accuracy and convergence}

A variety of tests were performed to determine both accuracy and the effect
on convergence rates of the new method. The 
RMG\cite{PhysRevB.54.14362,PhysRevB.76.085108} code developed at NCSU was
used for all tests. The first system consists of a 512-atom
diamond cell with a Vanderbilt\cite{PhysRevB.41.7892} ultrasoft
pseudopotential used to represent the ionic cores. Test runs
using submatrix widths ranging from 0.1 to 0.3 show that a width of 0.3
produces results identical to those obtained using standard full
diagonalization. While the size of this system is typical of production SCF
calculations, it is a perfect crystal with highly degenerate eigenvalues. Many
systems of interest are disordered and we expect that they should actually be 
easier to handle computationally because the degeneracies will normally be
broken in such systems.

\begin{figure}[ht]
\includegraphics[scale=0.33,angle=270]{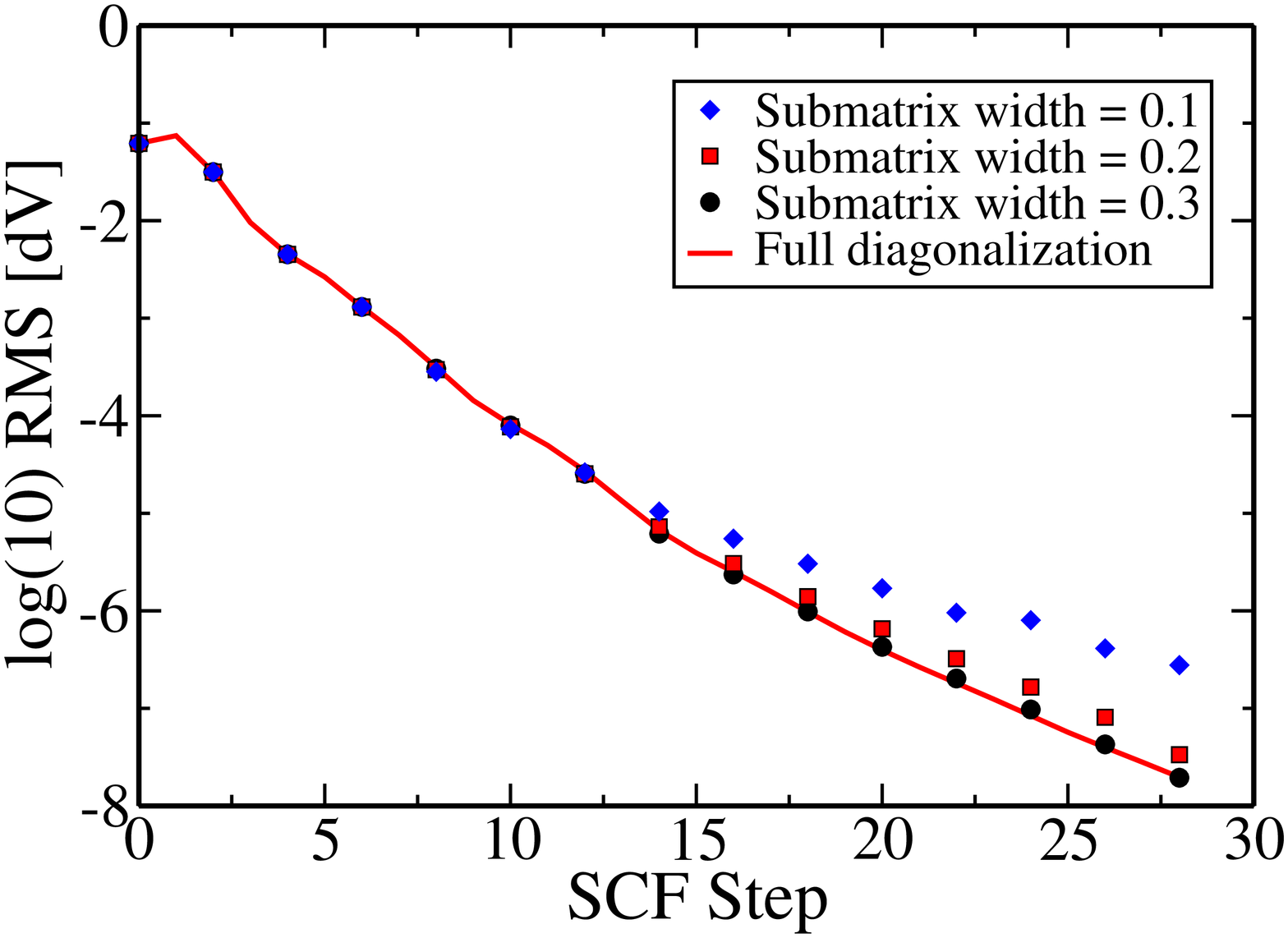}
\caption{Convergence rates for a 512 atom diamond cell for 
various submatrix widths.}
\label{diamond512_convergence}
\end{figure}

As an example of a disordered system we selected a cluster of 13 C60
molecules in a non-equilibrium FCC structure. This system again exhibits
identical results between standard full diagonalization and the new method
once the submatrix width reachs 0.3. 

\begin{figure}[ht]
\includegraphics[scale=0.33,angle=270]{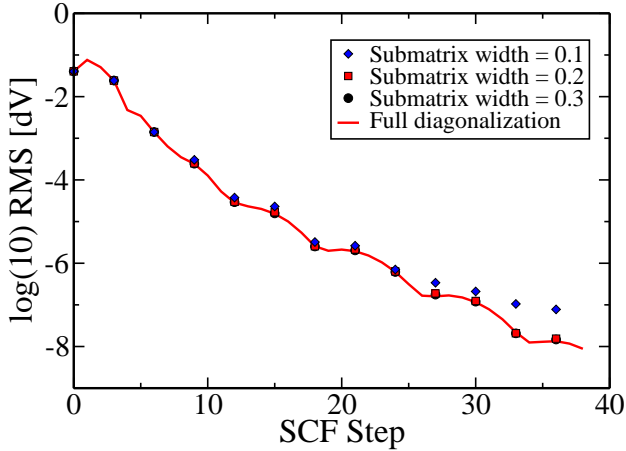}
\caption{Convergence rates for a cluster of 13 C60 molecules
(780 atoms total) for various submatrix widths.}
\label{c60cluster_convergence}
\end{figure}

Since diamond and c60 both have bandgaps, bulk copper is selected as an
example of a metallic system. Orbital occupations near the Fermi level are
assigned using a Fermi-Dirac distribution with $\mu = 0.04$ eV. In this
case the overall convergence rate was slower, as expected for a metal, but
the submatrix width required in order for the PFSM results to match those
from full diagonalization is reduced from 0.3 to 0.2. A similar reduction
of the required submatrix width has also been observed with other metallic
systems, which we attribute to the breaking of degeneracies due to the
Fermi-Dirac distribution of the orbital occupations.

\begin{figure}[ht]
\includegraphics[scale=0.33,angle=270]{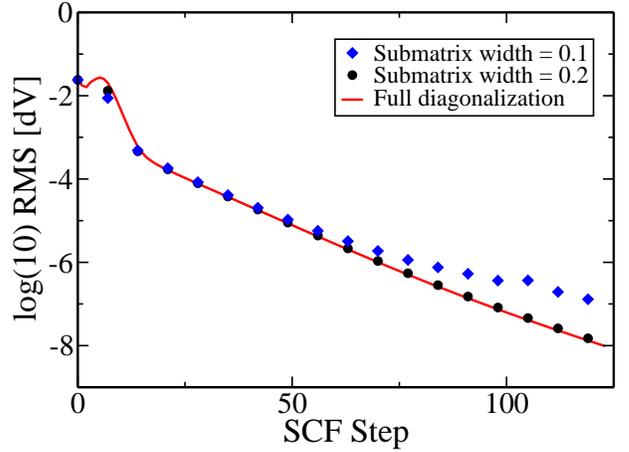}
\caption{Convergence rates for a 256 atom copper cell for various
submatrix widths.}
\label{copper256_convergence}
\end{figure}

\section{Performance and parallelization}

The PFSM method can use different types of eigensolvers for the submatrices
and RMG currently supports LAPACK and MAGMA. Improvements in execution time
depend on problem size and type, and the architecture of the target
platform. As the bulk of the execution time for the RMG code is normally
spent in the multigrid preconditioner, which exhibits $O(N^2)$ scaling,
and the eigensolver that exhibits $O(N^3)$ scaling, we expect that the
overall speedup will increase as $N$ increases and this is indeed
observed. The results presented in the following tests use the Bluewaters
system, a Cray XK7 located at NCSA, where each processing node consists of
a multi-core Opteron CPU and a single Nvidia K20X GPU. The nodes are connected
via a 3-D torus, with each link providing up to 9.6 GBytes/sec of 
bidirectional bandwidth. The current version of PFSM used in the RMG code
is targeted  towards systems where $N \le 10000$. This limit is not specific
to the algorithm, but rather is a consequence of the current implementation
that assigns each submatrix to a single processing node. As this problem
size is quite common in electronic structure calculations, the method is
widely applicable.

Several test systems were selected to illustrate the potential performance
gains. The timings depend on the submatrix width, which is always selected
so that convergence rates and final results are essentially identical for
PFSM and full diagonalization. Results for the 256 atom copper cell that were
used in the convergence tests illustrated in Fig.(\ref{copper256_convergence})
are presented in Table \ref{copper256_performance}. The improvement in the
eigensolver is a factor of 2.28 for MAGMA and 14.16 for LAPACK. The overall
speedup factors are 1.08 and 2.72 respectively.

\newcommand{\specialcell}[2][c]{%
  \begin{tabular}[#1]{@{}c@{}}#2\end{tabular}}

\begingroup
\squeezetable
\begin{table}[ht]
\centering
\begin{tabular}{|c|c|c|c|c|c|l|}
\hline
 & MAGMA & ScaLapack & LAPACK & \specialcell{MAGMA \\ PFSM} & \specialcell{LAPACK \\ PFSM} \\
\hline
Total time & 5.93 &  6.52 & 15.36 & 5.46 & 5.64 \\
\hline
Eigensolver & 1.23 & 1.83 & 10.34 & 0.54 & 0.73 \\
\hline
\end{tabular}
\caption{\label{copper256_performance}Time per SCF step (seconds) for 256
atom copper cell with 256 unoccupied orbitals, N=1664, using
64 Cray XK7 nodes.}
\end{table}
\endgroup

Results for the next system are presented in Table \ref{si216results}. These
results were generated for a 216 atom silicon vacancy with approximately 40
unoccupied states per atom. This provides a good example of the performance
improvements possible for systems with a large number of unoccupied states,
which are common in GW calculations. In contrast to the previous result,
where the MAGMA eigensolver was faster than ScaLapack, we now see better
performance from ScaLapack. This is not surprising since the MAGMA result is
being generated by a single GPU which is saturated for this value of $N$.
The Eigensolver portion of the MAGMA-PFSM result is in turn more than 5 times
faster than the pure MAGMA result and 2.6 times faster than the ScaLapack
result. We also note that the eigensolver results for LAPACK-PFSM are more
than 20 times faster than the LAPACK result and indeed are nearly twice as
fast as ScaLapack.

\begingroup
\squeezetable
\begin{table}[ht]
\centering
\begin{tabular}{|c|c|c|c|c|c|l|}
\hline
 & MAGMA & ScaLapack & LAPACK & \specialcell{MAGMA \\ PFSM} & \specialcell{LAPACK \\ PFSM} \\
\hline
Total time & 69.07 & 43.39 & 317.23 & 25.30 & 30.40 \\
\hline
Eigensolver & 52.04 & 26.75 & 296.67 & 10.36 & 14.23 \\
\hline
\end{tabular}
\caption{\label{si216results}Time per SCF step (seconds) for 216 atom
silicon vacancy 40 unoccupied orbitals/atom, N=9136, using
64 Cray XK7 nodes.}
\end{table}
\endgroup

The last test case presented in Table \ref{al4000results} is a 4000 atom
aluminum cell running on 512 Cray XK7 nodes. The number of unoccupied states
per atom is less than 1, which leads to a value of $N$ that is only slightly
larger than in the previous system, even though the cell volume is 14.9 times
larger. The Eigensolver portion of the ScaLapack calculation is 2.35 times
faster than the MAGMA result, while the MAGMA-PFSM result is 6.15 times faster
than the pure MAGMA result.

\begingroup
\squeezetable
\begin{table}[ht]
\centering
\begin{tabular}{|c|c|c|c|c|c|l|}
\hline
 & MAGMA & ScaLapack & LAPACK & \specialcell{MAGMA \\ PFSM} & \specialcell{LAPACK \\ PFSM} \\
\hline
Total time & 64.75 & 40.28 & 351.16 & 25.27 & 35.78 \\
\hline
Eigensolver & 45.25 & 19.27 & 323.17 & 7.35 & 12.40 \\
\hline
\end{tabular}
\caption{\label{al4000results}Time per SCF step (seconds) for 4000 atom
aluminum supercell, N=9216, using 512 Cray XK7 nodes.}
\end{table}
\endgroup

\section{Implementation}
The matrix elements of Eq(\ref{eq:kohnsham}) are generated as shown below.

\begin{equation}
\label{eq:kohnshammatrix}
\bm \left\langle \psi_i|H_{ks}|\psi_j \right\rangle = 
\lambda_i \left\langle \psi_i|\psi_j \right\rangle \;\;\; i,j=1,N
\end{equation}

This produces a generalized eigenvalue problem.

\begin{equation}
\label{eq:generalizedeigenproblem}
\bm A\bm x_i = \lambda_i \bm B \bm  x_i \;\;\; i=1,N
\end{equation}

Which is normally solved using a standard LAPACK or ScaLapack routine such
as DSYGVD or PDSYGVD. These routines convert the generalized form into the
standard form of Eq.(\ref{eq:eigenproblem}) before solving for the eigenpairs.
We use a similar approach in the PFSM but take advantage of specific
characteristics of the problem to do the conversion efficiently on massively
parallel computer architectures. Conversion to standard form is equivalent to
multiplying both sides of Eq.(\ref{eq:generalizedeigenproblem}) by 
$\bm B^{-1}$. The explicit computation of $\bm B^{-1}$ is not required
however and since $\bm B$ is diagonally dominant the identity serves as a 
good initial approximation, ($\bm B \ne \bm I$ at convergence due to ultrasoft
pseudopotentials\cite{PhysRevB.76.085108}). The initial guess may be improved using
an iterative process. Let

\begin{equation}
\label{eq:convert1}
\bm Z_{n+1} = (\bm I - \bm D^{-1} \bm B) \bm Z_n + \bm D^{-1} \bm A \bm x
\end{equation}

where $\bm D$ is the diagonal of $\bm B$. Then at convergence

\begin{equation}
\label{eq:convert2}
\bm Z = \bm Z - \bm D^{-1} \bm B \bm Z + \bm D^{-1} \bm A \bm x
\end{equation}

therefore $\bm B \bm Z = \bm A \bm x$ and $\bm Z = \bm B^{-1} \bm A$. This
iteration has the desirable property that the individual columns of Z can be
computed independently. Given $P$ computational nodes, we assign $N / P$
columns of $\bm Z$ to each node and achieve excellent scalability. After the
completion of this step an MPI\_Allgatherv call is used to ensure that all
processing nodes have a copy of $\bm Z$ that is now used in 
Eq(\ref{eq:eigenproblem}).

The matrix $\bm Z$ obtained in Eq.(\ref{eq:convert2}) is then decomposed over
processing nodes in the manner illustrated in Fig.(\ref{submatrices}). While
the current implementation only uses MAGMA or LAPACK running on a single node
to solve each submatrix, this is not an actual limitation of the method.
Groups of processing nodes running ScaLapack or some other form of
distributed eigensolver could also be used to solve the submatrices, work
is under way to implement this. We expect that this will only be advantageous
when the value of $N$ exceeds 15,000, because for submatrix sizes
of $M \le 3000$
(corresponding to full matrix sizes of 10000 to 15000), MAGMA running on a
single node is the fastest option for computing the submatrix solutions on
the Cray XK7. One consideration when using MAGMA is that the submatrix size
is constrained by the amount of GPU memory available on a single node. For
the XK7 this is 6GB and a considerable fraction of that memory is reserved
for other datastructures. Consequently, the present RMG implementation is
currently restricted to values of $N \le 10000$. We expect to expand this
to values of
$N \approx 35,000$ for the XK7 via improved memory management.
Future architectures, such as the Summit system scheduled to be installed at
ORNL in 2017, will have a larger coherent memory space accessible to the GPU
accelerators (up to 512 GBytes/node), and should be able to handle
much larger systems.

\section{Summary}
We have developed a new method, referred to as PFSM, for computing the
eigenpairs of matrices of the type typically found in \emph{ab-initio}
electronic structure calculations. PFMS exploits specific characteristics of
the problem, and the available hardware architectures to achieve up to an
order of magnitude improvement in the eigensolver execution time. The
improved speed can be obtained with no sacrifice in accuracy or convergence
rates, and is particularly well suited to metallic systems. The source code
for PFSM is currently available in the latest stable release of RMG, which
may be obtained at http:://rmgdft.sourceforge.net. It is integrated into
the main code base and work is underway to move it into a separate library
with a calling interface similar to the standard eigensolvers. This should
make it easy to incorporate the method into other electronic structure
codes with minimal work.

\section{Acknowledgements}
We gratefully acknowledge discussions with Drs. Miroslav Hodak and Wenchang
Lu. EB was supported by NSF grant ACI-1339844, CTK by NSF ACI-1339844 and
DMS-1406349, and JB by DOE DE-FG02-98ER45685.

\bibliography{Newdiag}

\end{document}